\documentclass[
superscriptaddress,
 amsmath,amssymb,
 aps,
prd,
twocolumn,
floatfix
]{revtex4-2}

\usepackage{graphicx}
\usepackage{dcolumn}
\usepackage{bm}
\usepackage{url}
\usepackage{xcolor} 
\usepackage{physics} 
\usepackage{setspace} 
 \usepackage{float}

\newcolumntype{M}[1]{>{\centering\arraybackslash}m{#1}}
\newcolumntype{N}{@{}m{0pt}@{}}

\def\prn#1{{\left(#1\right)}}

\begin{document}

\title{GPS constellation search for exotic physics messengers coincident with the binary neutron star merger GW170817
}

\author{Arko P. Sen}
\affiliation{Department of Physics, University of Nevada, Reno, Nevada 89557, USA}

\author{Geoffrey Blewitt}
\affiliation{Department of Physics, University of Nevada, Reno, Nevada 89557, USA}
\affiliation{Nevada Bureau of Mines and Geology, University of Nevada, Reno, Nevada 89557, USA}

\author{Andrey Sarantsev}
 \affiliation{Department of Mathematics and Statistics, University of Nevada, Reno, Nevada 89557, USA}

\author{Paul Ries}
 \affiliation{Jet Propulsion Laboratory, California Institute of Technology, Pasadena, 
 California 91109, USA}

\author{Andrei Derevianko }
\email{andrei@unr.edu}
\affiliation{Department of Physics, University of Nevada, Reno, Nevada 89557, USA}

\date{\today}

\begin{abstract}
The Global Positioning System (GPS) includes a continuously operating, planet-scale network of atomic clocks that, beyond navigation and time dissemination, enables precision tests of fundamental physics.
Here we use GPS carrier phase archival data to perform a retrospective search for exotic low-mass fields (ELFs) that might be emitted by the binary neutron-star merger GW170817, complementing gravitational wave
and electromagnetic modalities
in multi-messenger astronomy.
Such ultra-relativistic fields would imprint a dispersive, anti-chirp signature in clock-frequency time series,
delayed with respect to the LIGO-Virgo gravitational wave detection.
We construct network-median pseudo-frequency data from eighteen Rb satellite clocks referenced to a terrestrial hydrogen maser and conduct a template-bank search spanning ELF pulse duration, arrival delay, and characteristic frequency. No statistically significant signal is observed after accounting for noise statistics and template-bank trials. We derive 95\% confidence-level lower bounds on the interaction energy scale $\Lambda_\alpha$ of quadratic couplings driving variations in electromagnetic fine-structure constant. These limits improve upon existing astrophysical and gravity-test constraints across the ELF-energy range $\approx10^{-18}$--$10^{-14}\,\mathrm{eV}$. This demonstrates that mature global satellite-clock networks provide an observational capability for retrospective, multi-messenger searches for new physics using decades of archival timing data.
\end{abstract}

\maketitle


The Global Positioning System (GPS) constellation consists of  ${\approx 32}$ satellites~\cite{Hoffman}. By broadcasting time-stamped signals referenced to onboard atomic clocks together with orbital ephemerides, GPS enables receivers to determine their position and clock offset with accuracy at the meter level and nanosecond level, respectively.  Beyond navigation, GPS serves as a cornerstone for high-precision geodesy, a critical tool for geophysics~\cite{BLEWITT2015307}. For such applications, the analysis of dual-frequency carrier-phase data produces substantially improved accuracy at the few-mm level for positioning and ${\sim 0.01\,\mathrm{ns}}$ for timing~\cite{BERTIGER2020469}. Building on these high-end capabilities, GPS has become a mature platform for fundamental physics. 
For example, long-term GPS clock data may constrain violations of Einstein's equivalence principle and of local Lorentz and position invariance~\cite{KenMoh2012-PlankGPS,FatWuPon2023-GPSRedShift}.
More recently, correlated timing anomalies across the GPS constellation have been exploited to search for exotic physics, such as transient encounters with clumpy dark matter~\cite{DerPos14,Roberts2017-GPS-DM}. GPS can thus be used as a laboratory for precision tests of fundamental laws on planetary scales.

Here we use the network of GPS atomic clocks as a $\sim 50,000$-km aperture observatory for multi-messenger astronomy in exotic physics modality.  Multi-messenger astronomy is a coordinated observation of different classes of signals originating from the same astrophysical event~\cite{Abbott_multimessenger_2017}. An event of particular interest to us is the binary neutron star (BNS) merger GW170817 that was detected in both gravitational wave (GW)~\cite{abbott2017gw170817} and multi-wavelength electromagnetic radiation~\cite{Coulter2017,Hallinan2017,Soares-Santos2017,Troja2017,Margutti2018,Mooley2018,Ruan2018} modalities. 
The messengers came from a host galaxy 40 megaparsecs ($R \approx$ 130 million light years) away. Ref.~\cite{dailey2020ELF.Concept} hypothesizes that such events might 
be accompanied by an emission of exotic fields that are detectable using GPS clocks. To test this hypothesis we use GPS archival carrier-phase data in a window surrounding the date of the event Aug.\ 17, 2017.   
{Unlike dedicated laboratory experiments, GPS provides a continuously operating, globally distributed clock network with decades-long archival data, enabling retrospective multi-messenger searches that cannot be replicated by purpose-built detectors.} 

\begin{figure*}
\center
\includegraphics[width=0.95\textwidth]{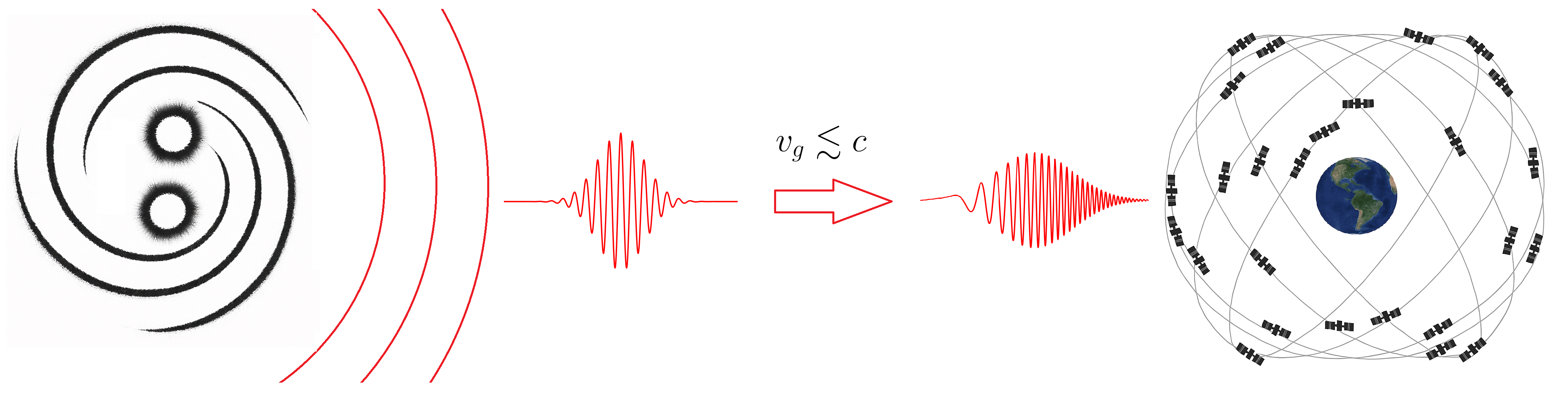}
\caption{A burst of exotic low-mass fields (ELFs) propagates with the group velocity $v_g \lesssim c$ lagging behind the gravitational wave burst traveling at speed of light, $c$. The arriving ELF wave-packet exhibits  a characteristic  frequency anti-chirp~\cite{dailey2020ELF.Concept}, because of the faster propagation of the more energetic ELF components. GPS constellation atomic clocks  serve as a distributed quantum sensor. The GPS constellation image  is adapted from the GPS.gov public domain image.
}
\label{Fig:ELF}
\end{figure*}

Fig.~\ref{Fig:ELF} illustrates the idea of multi-messenger astronomy involving exotic low-mass fields (ELFs). In this scenario, a merger of massive compact bodies may produce both GWs and a burst of ELF radiation. Given ELF's small but finite mass $m$, an ELF pulse propagates at a group velocity $v_g\lesssim c$, where $c$ is the speed of light. Therefore, there is a time delay $\delta t$ between the GW trigger and ELF pulse arrival. For reasonable $\delta t$, ELFs must be ultrarelativistic considering the vast distances traversed.  
Dispersion of the ELF pulse causes it to spread out  with higher-frequency components arriving earlier. This results in a characteristic anti-chirp signature (a downward frequency sweep over time) that might be detectable by the GPS atomic clock networks~\cite{dailey2020ELF.Concept,Derevianko2023-Moriond.ELF}.

We compare GPS clock data taken on the date of the BNS GW detection, Aug.\ 17, 2017, with the signals driven by the ELF waveforms shown in Fig.~\ref{Fig:ELF}.
This waveform is parameterized as~\cite{dailey2020ELF.Concept} 
\begin{align}
\phi(t) \approx  \frac{A_0}{R} \exp\left[-\frac{(t-t_s)^2}{2 \tau^2} \right] \cos\theta(t)\,,
\label{Eq:Waveform}
\end{align}
where $t_s = t_\mathrm{GW} +  \delta t$ is the moment the center of the pulse arrives at the sensor and $R$ is the distance to the progenitor, so $t_\mathrm{GW} = R/c$. The pulse amplitude $A_0={\sqrt{c\Delta E/(2\pi^{3/2} \tau)}}/{\omega_0}$. Here $\omega_0$ is the ELF pulse central frequency and $\Delta E$ is the energy released in the ELF channel.
The duration $\tau$ of the arriving pulse is related to the initial pulse duration $\tau_0$ as  $\tau =\tau_0\sqrt{1+\xi^2}$,
where $\xi={\Delta v_g t_s}/{v_g \tau_0}$ is determined by the spread in the ELF group velocities $\Delta v_g$. The phase of the ELF pulse is given by
\begin{eqnarray}
 \theta(t)& = & \omega_0 \left(t-t_\mathrm{GW}\right) -\frac{1}{2\tau^2}\frac{\xi t}{t_s}\left(t-t_s\right)^2  \nonumber\\
 &+&\frac{1}{2}\tan^{-1}\left(\frac{\xi t}{t_s} \right)+\theta'\,,\label{Eq:phase1}   
\end{eqnarray}
where $\theta'$ is an unknown phase offset accumulated by the ELF pulse as it traverses the interstellar medium. The waveform~\eqref{Eq:Waveform} can be derived by noticing~\cite{Derevianko2023-Moriond.ELF} that for ultrarelativistic scalars, the mass-induced index of refraction is $n\left(  \omega\right) \approx 1-{m^2 c^4}/{2\hbar^2\omega^{2}}$ and by applying results from electromagnetic wave propagation theory~\cite{JacksonEM}. For example, $v_g =c/\left(  n+\omega dn/d\omega\right)$.  Overall, for a given progenitor, the ELF waveform is uniquely determined by specifying $\tau_0$, $\delta t$ and $\omega_0$. See Supplemental Material  for further details. 

GPS relies on satellite and terrestrial atomic clocks. Here we use $^{87}$Rb satellite clocks, given their greater prevalence and the well-characterized, predominantly white nature of their frequency noise~\cite{DanzeyRiley,Roberts2018a}.
GPS satellites broadcast microwave signals with a carrier phase that is commensurate with the on-board atomic clock system, through a phase-locked loop frequency synthesizer~\cite{Hoffman}. These carrier phases are measured by a global terrestrial network of GPS receivers.  GPS data processing techniques are then used to model and invert the carrier phase data to estimate many network parameters, including the GPS clock biases for each satellite and receiver relative to a selected reference clock~\cite{Bertiger2010}. Our selected reference clock is a hydrogen maser at receiver station KOKV, situated in Kauai, Hawaii~\cite{ITRF_KOKV}.

The Jet Propulsion Laboratory routinely analyzes carrier phase data every 30 seconds from the global GPS network using GipsyX software~\cite{BERTIGER2020469}. To improve time resolution for this experiment, we reprocessed the archival carrier phase data at the higher rate of 1\,Hz from a global 30-station network, to produce clock data every 1\,s.  Our analysis spans the day of the GW170817 BNS merger and the previous two days. Data processing includes modeling the time of flight of the carrier signal traveling at the phase velocity of light. Modeled effects include~\cite{BLEWITT2015307} GPS orbit parameters, solar radiation pressure, station positions, Earth rotation, relativistic time dilation, gravitational red shift, Shapiro delay, phase rotation, antenna phase center variation calibrations, atmospheric refraction, and tidal motion. 
Ionospheric refraction is mitigated by a combination of carrier phases measured at two different frequencies, together with higher-order corrections~\cite{Bertiger2010}. Integer ambiguities in the carrier phase data are tightly constrained~\cite{Blewitt_1989, Bertiger2010}. Further details on data processing are presented in Supplemental Material.

The 1-Hz sampling rate and the data availability limit our search to ELF energies in the interval $ 10^{-19}\, \mathrm{eV} \lesssim \varepsilon_0=\hbar \omega_0 \lesssim  10^{-14} \,  \mathrm{eV}$, Ref.~\cite{SenPfeRies2024-GPS.ELF}. The GPS constellation has a diameter $D_{\rm{GPS}} \approx 5\times 10^4$~km. Ultrarelativistic ELFs travel across the constellation in $\approx0.2\,\mathrm{s}$, much shorter than our sampling time interval. Moreover, all network clocks would experience approximately the same field value, when the ELF wavelengths $\lambda_\mathrm{ELF} \approx { 2\pi \hbar  c}/{\varepsilon_0} \gg D_{\rm{GPS}}$. For the specified range of ELF energies, {$10^{5} \lesssim \lambda_\mathrm{ELF} \lesssim 10^{10}$}~km. Then the entire GPS constellation can be treated as a composite point-like quantum sensor in our search. 
{Such composite sensor helps in vetoing  signals that do not affect all $N_c$ clocks and gains $\sqrt{N_c}$ in sensitivity.}
Note that in this regime, ELF signal detection depends on the reference clock having a different sensitivity to the variation of fundamental constants than the satellite Rb clocks, hence our selection of an H-maser reference clock.

Ref.~\cite{dailey2020ELF.Concept} demonstrated that the GPS clocks are sensitive to hypothetical GW170817 BNS-emitted ELFs $\phi(t)$ interacting with atoms through  the phenomenological Lagrangian
\begin{align}
   \mathcal{L} = \left({-\sum_f\Gamma_f  m_f c^2\bar{\psi}_f\psi_f + \frac{\Gamma_\alpha}{4} F_{\mu\nu}F^{\mu\nu} }\right)\hbar c\phi(t)^2\,.\label{Eq:Lagrangian2}
\end{align}
Here $\psi_f$ are the fermion fields of masses $m_f$ and $F_{\mu \nu}$ is the electromagnetic Faraday tensor. $\Gamma_X$ are unknown coupling strengths. Such interactions were studied in  multiple contexts~\cite{OlivePospelov,DerPos14, BouSorYu2023-quadratic} ranging from supernovae emission to dark matter and Big Bang nucleosynthesis. Importantly, $\mathcal{L}$ leads to variations of fundamental constants that affect atomic energy levels and, thereby, the measured atomic clock frequencies~\cite{DerPos14}. 

The fractional clock frequency excursions due to  interactions~\eqref{Eq:Lagrangian2} can be expressed as
\begin{align}
    s(t) =\Gamma_{\mathrm{eff}}\hbar c\phi(t)^2 \,.
    \label{Eq:ELF-sig}
\end{align}
For the $^{87}$Rb satellite clock - terrestrial H-maser pairwise comparisons carried out in our search,  the effective coupling constant  is 
$\Gamma_{\mathrm{eff}}=0.34\,\Gamma_{\alpha}+0.081\,\Gamma_{m_q}$~\cite{SenPfeRies2024-GPS.ELF}, where $\Gamma_{m_q}$ is the quark mass coupling strength and $\Gamma_{\alpha}$ parametrizes the sensitivity to variations in fine structure constant $\alpha$. Following the literature, we parametrize the interaction strength by the energy scale, e.g., $\Lambda_{\alpha}\equiv 1/\sqrt{|\Gamma_{\alpha}|}$. 

\begin{figure}[ht!]
    \centering  \includegraphics[width=1.0\columnwidth]{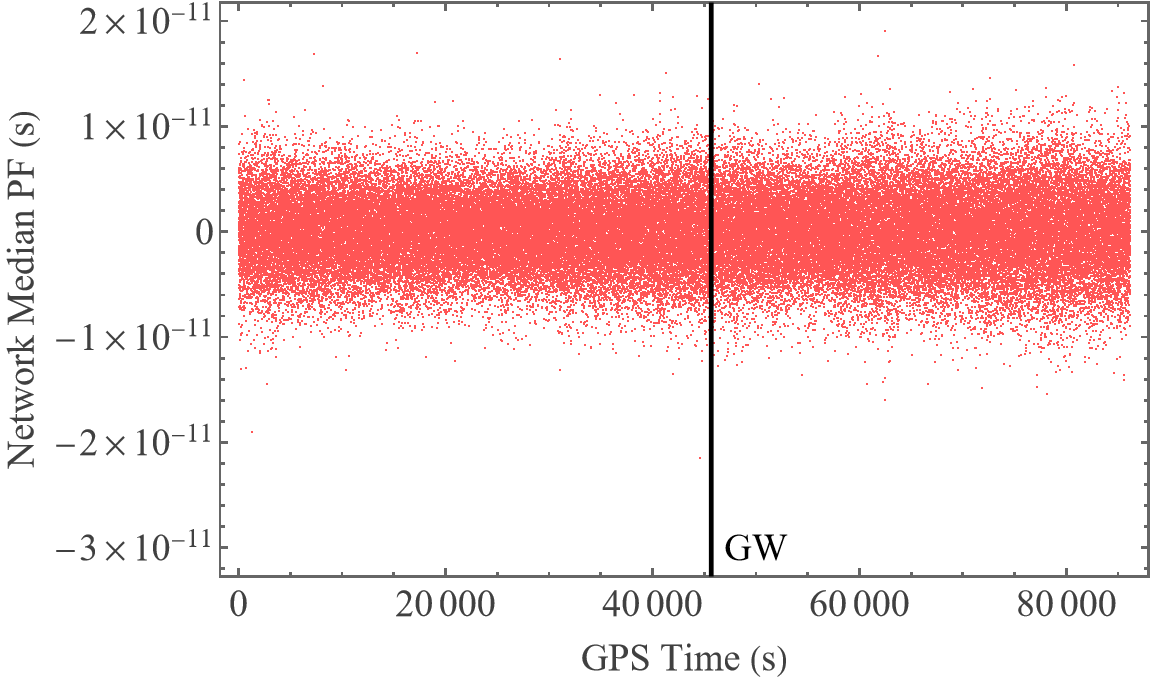}
    \caption{ GPS network median clock pseudo-frequency (PF) for Aug.\ 17, 2017. LIGO-Virgo gravitational wave (GW) trigger~\cite{Abbott_multimessenger_2017} (black vertical line) is at $45,682\,\mathrm{s}$ from the start of the GPS day of Aug.\ 17, 2017. Network median is taken over $18$ Rb clocks of the GPS network. The difference between GPS and UTC time scales here was $18\,\mathrm{s}$. Visually there is no obvious change in noise behavior after the GW trigger. 
    }\label{Fig:Net_Med_Plot}
 \end{figure}

Our data sets contain clock biases (time offsets) of $18$~$\mathrm{Rb}$ satellite clocks relative to the KOKV H-maser. The Rb clock bias time series are dominated by a constant frequency drift plus random-walk noise. To assist statistical analysis, we whiten the data by differencing the clock biases~\cite{Roberts2017-GPS-DM}. We refer to such data as clock pseudo-frequencies (PF). Since $\lambda_\mathrm{ELF} \gg D_\mathrm{GPS}$,
all the clocks would be affected by the ELF pulse identically, thus for every second we take PF median over the network. We use the median to capture signals that significantly affect the majority of satellite clocks, while ignoring signals that do not.
The resulting time series for the date of the event is shown in Fig.~\ref{Fig:Net_Med_Plot}. We form similar PF series for two days prior to the event. We take into account that the orbital geometry of the constellation repeats relative to an observer on Earth with the orbit repeat time in the Earth-fixed reference frame of 1~day minus 245~s = $86,155\,\mathrm{s}$ for all satellites~\cite{Agnew-repeat}. We included such time-shifting to suppress repeating common-mode errors when comparing PFs over multiple days~\cite{Larson-repeat}.

{To visualize the noise properties of the network-median PF data, we divide each time series into equal-length windows and compute, within each window, the standard deviation $\sigma_w$ relative to the local mean. Fig.~\ref{Fig:STDEV} compares $\sigma_w$ for Aug.\ 16 and 17, 2017 (Aug.\ 15 is in Supplemental Material). The temporal structure on the GW-event day closely follows that of the preceding days, consistent with the daily repeating common-mode behavior of the constellation. Local variations in $\sigma_w$ are present on all days and remain within the variability expected from the two-day period prior to the merger and do not exhibit any obvious structure associated with an ELF signal. These diagnostics do not reveal any features that would motivate a standalone detection claim.}

\begin{figure}[h!]
    \centering  \includegraphics[width=1.0\columnwidth]{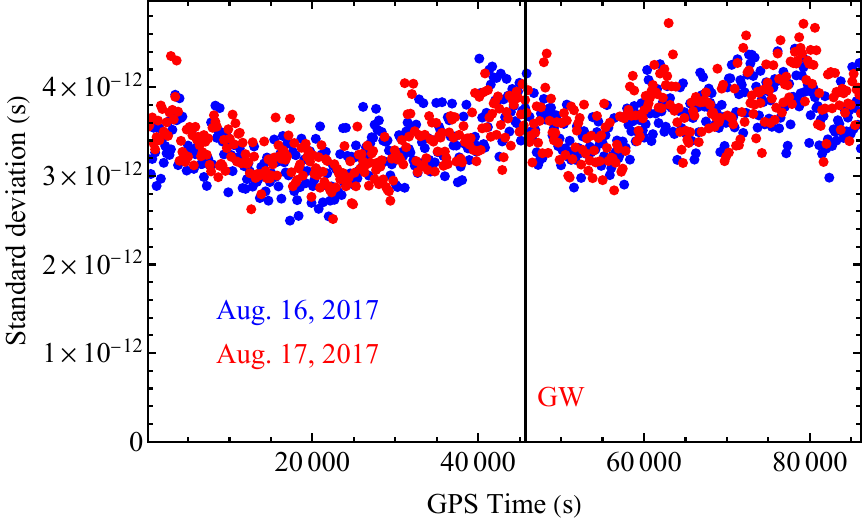}
    \caption{Standard deviations $\sigma_w$ of GPS network median pseudo-frequencies for Aug.\ 16 and 17, 2017. GW trigger on Aug.\ 17, 2017, is shown with the black vertical line.  Standard deviations were computed over 173-s wide windows. The GW-event day shows variance behavior consistent with the previous day control period, with no structure indicative of an ELF signal.
    }
    \label{Fig:STDEV}
\end{figure}

Given that data in Figs.~\ref{Fig:Net_Med_Plot} and \ref{Fig:STDEV} do not reveal obvious ELF signatures, we apply more rigorous strategies to search for the putative ELF signals. 
The expected ELF signal~\eqref{Eq:ELF-sig} is a known time-dependent template with an overall amplitude 
$h= \Gamma_{\mathrm{eff}} \hbar c A_0^2/R^2$
that encodes the underlying physics and whose Gaussian envelope and phase evolution, c.f. Eqs.~\eqref{Eq:Waveform} and \eqref{Eq:phase1}, account for dispersion and timing relative to the LIGO trigger. The network median PF data of the two days prior to the event are shown to have an approximate Gaussian distribution according to the Anderson-Darling test~\cite{DAgostinoStephens1986,AndersonDarling1954}. 
Then, the amplitude estimator $\hat{h}$ is obtained by correlating the data with the template, with its variance $\hat{\sigma}^2_h$ set by the noise level and template norm. This naturally defines a signal-to-noise–ratio (SNR) statistic as $|\hat{h}|/\hat{\sigma}_h$.

The ELF search is carried out using a bank of templates parameterized by triples $(\tau_0,\delta t,\omega_0)$ specifying the initial pulse width, arrival delay, and central frequency. The unknown phase $\theta'$ in Eq.~\eqref{Eq:phase1} is effectively treated as a marginalization parameter; for a fixed triple, we maximize SNR over $\theta'$. 
Physical and sampling constraints restrict $(\tau_0,\delta t,\omega_0)$ combinations, ensuring that waveforms are well resolved in time and that the underlying approximations remain valid. Scanning this space yields a large $\sim 10^{13}$ number of templates, which collectively form the template bank. 

Our search methodology follows a frequentist hypothesis-testing framework~\cite{RomanoCornish2017}. Under the null hypothesis of noise-only data, the SNR follows a known distribution, enabling detection thresholds to be set at fixed false-alarm probabilities. Because the large number of templates greatly increases the chance of  high-SNR fluctuations, the detection threshold must be raised accordingly~\cite{Daykin2021}; this threshold ($\approx 8$) is substantially higher than the conventional single-template value of $3$. A fully exhaustive search is computationally prohibitive, so our search proceeds in stages: an initial sparse scan through the bank finds no significant candidates. 
{Next, a denser search focuses on short-duration pulses, as these dominate the background variability characterized in Fig.~\ref{Fig:STDEV}.
None yield SNRs exceeding the template-bank–adjusted detection threshold. We therefore find no evidence for an ELF signal.}

In the absence of a statistically significant ELF signature, we set upper limits using a frequentist construction based on the SNR distribution. Adopting a 95\% confidence level, this procedure maps each observed SNR to a corresponding upper bound on the true SNR, which in turn yields a 95\% upper limit on the inferred signal amplitude $h$ and thus on the coupling constant $\Gamma_\mathrm{eff}$. 
Notice that the sign of the coupling $\Gamma_\mathrm{eff}$ plays an important physical role~\cite{Stadnik2024-screeningNA,Derevianko2024-ELFreplyNA}. For positive couplings, interactions with ambient Standard Model matter increase the effective mass of the ELF field, leading to environmental screening that attenuates the signal at the detector. In contrast, negative couplings avoid this screening over a wide parameter range, allowing ELF waves to propagate freely and remain detectable~\cite{Derevianko2024-ELFreplyNA}. 

\begin{figure}[ht!]
    \centering  \includegraphics[width=1.0\columnwidth]{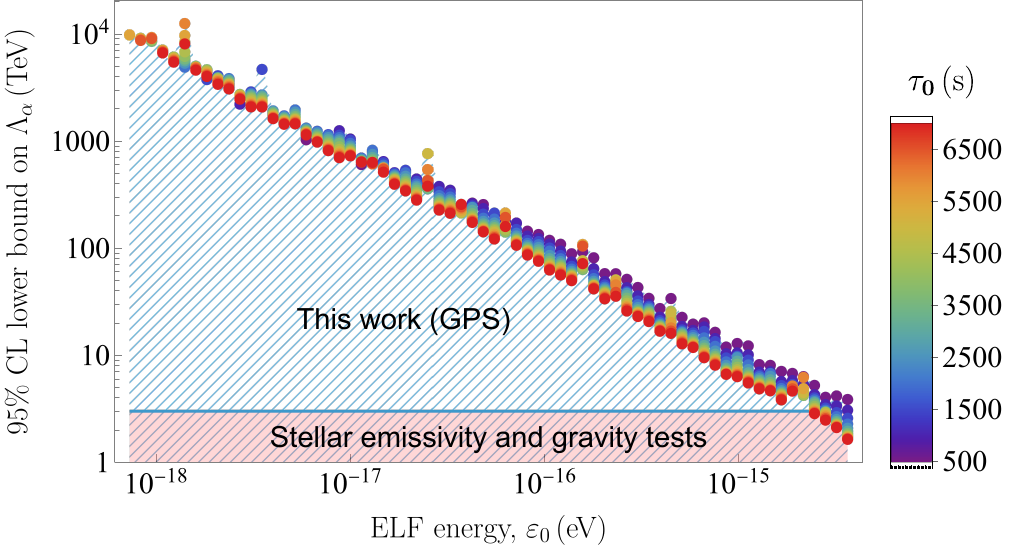}
    \caption{ Representative GPS constraints on ultralight scalar fields from GW170817.
Shown are 95\% confidence-level (CL) lower bounds on the energy scale $\Lambda_\alpha$
governing quadratic couplings of ultralight scalar fields causing variations in fine-structure constant. These are derived from the non-observation of an exotic-field signal temporally
correlated with the binary neutron star merger GW170817. Bounds are presented as a function of the scalar-field energy $\varepsilon_0$ for a representative
GW--ELF time delay $\delta t = 15{,}000~\mathrm{s}$ and initial pulse durations
$\tau_0$ spanning $500$--$7{,}000\,\mathrm{s}$. 
The blue hatched region is excluded by our work.
 Existing constraints from stellar
emissivity and precision tests of gravity are shown for comparison (pink shaded
region). The GPS-based limits extend well beyond all previous terrestrial and astrophysical bounds across much of the accessible parameter space, establishing
global satellite atomic-clock networks as the most sensitive probes of relativistic ultralight scalar transients in this energy range.
}\label{Fig:LimPlot2}
 \end{figure}

We present the limits on the interaction energy scale~$\Lambda_{\alpha}$, in the assumption that ELFs primarily cause variation in $\alpha$. The resulting exclusion regions in Fig.~\ref{Fig:LimPlot2}, presented as functions of the ELF energy $\varepsilon_0$ for a representative delay and various initial pulse durations, demonstrate that the GPS-based search probes energy scales that extend well beyond existing constraints $\Lambda_{\alpha}\lesssim 3\,\mathrm{TeV}$ from astrophysical observations and precision tests of gravity~\cite{OlivePospelov}. These GPS limits were placed using the fiducial value~\cite{dailey2020ELF.Concept} $\Delta E = 0.1 M_\odot c^2 $ of the energy released in the ELF channel by the BNS merger.
{Alternatively, taking prior bounds~\cite{OlivePospelov} on  $\Lambda_{\alpha}$
as a reference, the ELF non-observation limits the ELF channel energy $\Delta E$  to  $ \lesssim 2 \times 10^{-9} M_\odot c^2$ for $\varepsilon_0 = 10^{-18} \, \mathrm{eV}$. } An additional exclusion plot is provided in Appendix.

{
These results establish mature global satellite clock networks as observatories for physics beyond the Standard Model in the era of multi-messenger astronomy. The retrospective ELF search technique developed here can be extended in future studies by applying it to alternative LIGO events and the corresponding GPS clock-bias data from the days of other gravitational-wave detections. 
More broadly, our GPS-based approach, together with complementary searches for axion-like ELF interactions pursued by the GNOME magnetometer collaboration~\cite{khamis2025}, provides a framework for probing ELF transients across a wider class of field–matter couplings. 
Looking ahead, substantially enhanced sensitivity may be achieved through ELF searches with global networks of entangled atomic clocks~\cite{Komar2016} or by exploiting the strong sensitivity of $^{229}$Th nuclear clocks~\cite{FadBerFal2020-ThClockAlphaVar,Rellergert2010} to variations of fundamental constants.}

\section*{Acknowledgments}
This work was supported in part by U.S. National Science Foundation Grant PHY-2207546, by the Heising-Simons Foundation, and by NASA Grants 80NSSC22K0463 and 80NSSC25K7865. The GPS carrier phase data used for this work was obtained from the International GNSS Service.  
The research was partly carried out at the Jet Propulsion Laboratory, California Institute of Technology, under a contract with the National Aeronautics and Space Administration (80NM0018D0004).

\appendix

\section{GPS Atomic clock data preparation}\label{Sec:GPS-clocks}

\subsection{GPS carrier phase and atomic clock phase}
\label{Sec:GPS-data-processing:carrier-phase}

We refer the reader to a detailed review~\cite{SenPfeRies2024-GPS.ELF} of the GPS atomic clocks operation and their architecture relevant to our ELF search.
To summarize, in GPS satellite clocks, a quartz oscillator is locked to the atomic transition frequency $\nu_{\mathrm{clock}}$ via a servo loop, which in turn drives a frequency synthesizer to produce the digital clock phase signal. The servo-loop time constant determines the duration of the fastest perturbation that the clock can record~\cite{SenPfeRies2024-GPS.ELF}. Our data, sampled once per second, allow sufficient time for the local oscillator to track the Rb oscillator. 

The phase of each satellite signal is sampled at regular intervals by a global terrestrial network of GPS receiver stations.  In our case, we use data sampled every $1\,\mathrm{s}$.  Each GPS receiver measures the beat phase between the incoming carrier signal mixed with an internal replica signal, which is locked in phase to the receiver's internal quartz oscillator. Thus, the beat phase measures the difference between the reception time according to the receiver clock and the transmission time according to the observed GPS satellite clock.  

This defines ``carrier phase data," which contain the imprint of  both clocks of the satellite and receiver station associated with each paired data stream.  Note that the carrier phase data are not simply the clock data we wish to obtain, as they are dominated by the changing range between the satellite and station. To obtain the clock data requires carrier phase data processing.

\subsection{GPS carrier phase data processing}\label{Sec:GPS-data-processing}

Data processing techniques are then used to model and invert all the carrier phase data from a global network of stations to estimate many network parameters, including the GPS clock biases for each satellite and station relative to a selected reference clock~\cite{Bertiger2010}. All satellite and station clock biases are estimated every $1\,\mathrm{s}$, except for the selected reference clock bias which is fixed to zero, by definition.

The Jet Propulsion Laboratory (JPL) routinely analyzes carrier phase data every 30 seconds from the global GPS network using GipsyX software ~\cite{BERTIGER2020469}.  For the special purpose of this ELF search, we processed data at 1\ Hz spanning multiple days from an identical global 30-station network for two days prior the GW170817 BNS merger and also on the day of the event. Data processing includes modeling the time of flight of the carrier signal traveling at the phase velocity of light. Modeled effects include~\cite{BLEWITT2015307} GPS orbit parameters, solar radiation pressure, station positions, Earth rotation, relativistic time dilation, gravitational red shift, Shapiro delay, phase rotation, antenna phase center variation calibrations, atmospheric refraction, and tidal motion.  More details on data processing can be found in Ref.~\cite{SenPfeRies2024-GPS.ELF}.

Satellite orbits are all modeled in an Earth-Centered Inertial (ECI) frame, including gravitational and non-gravitational forces (such as radiation pressure).  Orbit state parameters are estimated every day ($\sim$~2~orbit periods), including initial satellite position and velocity, and solar radiation pressure biases and their stochastic variation. Station positions are estimated in a co-rotating Earth-Centered Earth-Fixed (ECEF) frame using models of solid Earth tides and ocean loading deformation. Using a model of Earth rotation, the station coordinates in the ECEF frame are transformed into the ECI frame, in which the vacuum speed of light can be considered constant. Biases in polar motion and Earth's angular velocity are estimated and applied to the frame transformation.  This is made possible by modeling the dynamics of the satellites in the inertial frame, while observations are made on the rotating Earth.

Each satellite clock has its frequency purposely offset to allow for the mean combined relativistic effects of time dilation and gravitational red shift to a terrestrial observer~\cite{BLEWITT2015307}. This is achieved in the satellite firmware by using the monitored value of the orbit's semi-major axis. However, GPS satellite orbits are not precisely circular, each having a monitored eccentricity $\lesssim 0.02$. Thus, during each orbital period, a satellite will undergo a variation in speed and gravitational potential, which in turn causes a periodic variation in the observed satellite clock bias of amplitude $\lesssim 50$~ns.  By convention in GPS data processing, the relativistic variation in satellite clock bias is modeled assuming the Earth has a spherically symmetric mass distribution, and can be written
$-2{\bf r \cdot v}/c^2$, where ${\bf r}$ is the satellite's geocentric position, and ${\bf v}$ is the satellite's geocentric velocity. This relation holds in both ECI and ECEF frames, and includes both time dilation due to motion and gravitational red shift, in the approximation of Keplerian orbits.

Residual subdaily periodic relativistic effects that remain in the estimated satellite clock biases are dominated by the effect of Earth's oblateness with an amplitude of $\sim 0.1$~ns at $\sim 6$ hour period ~\cite{Kouba-relativity}.  The effect on fractional frequency is $\sim 2 \times 10^{-14}$, which is negligible compared to the frequency noise observed at $> 10^{-12}$.  Apart from the conventional relativistic term, systematic effects on estimated GPS clock biases appear to be dominated by non-relativistic (e.g. thermal) effects at $\sim$~1~ns, which tend to repeat at orbit periods ~\cite{Senior-GPS-clocks}.  This corresponds to long-period fractional frequency variations at $\sim 10^{-13}$.  In our search for ELFs, we are more concerned with short-period systematic variations, which can be mitigated by exploiting common-mode repeating patterns between neighboring days.

The resulting data set used in our analysis is a 1-s time series of atomic clock biases from each of the GPS satellites that have $^{87}\text{Rb}$ clocks, relative to a reference clock.  The reference clock used for purposes of parameter inversion was the H-maser clock at station KOUR, situated at Kourou, French Guiana.  The reference clock for all the clock bias data can easily be changed by subtracting clock bias data from the newly selected reference station.  This works because the reference station clock is common-mode to all clock bias time series.  For our analysis, we determined that the clock frequency noise was significantly lower at station KOKV, situated in Kauai, Hawaii~\cite{ITRF_KOKV}.  Hence we subtracted KOKV clock bias time series from all other time series, which effectively converts the reference clock from KOUR to KOKV.  We checked that our analysis is not sensitive to the selection reference clocks provided they have low frequency noise (for example, station YELL situated at Yellowknife, Canada).
 
 We also have access to receiver clock biases; however, most GPS stations only use quartz oscillators without an external atomic frequency standard, and so are of limited use.  Other output data include satellite positions and velocities, and station positions, which could be used if observing geometry is of interest, for example, in situations where the astronomical source direction is of interest, such as a dark matter search~\cite{Roberts2017-GPS-DM}.

\subsection{Description of submitted data files}
\label{Sec:GPS-data-processing:files}

Files submitted as part of the Appendix include clock bias data.  These clock data are obtained at 1-second intervals by processing the 1-Hz carrier phase data from a global network of stations using the GipsyX software at JPL. The provided clock bias data files are in ``Time Dependent Parameter'' (TDP) format. A description of the TDP format is provided in submitted files.

Clock bias data include the following satellites, which have Rb clocks with complete data sets over all three days: GPS36, GPS41, GPS43, GPS44, GPS46, GPS48, GPS55,
GPS60, GPS61, GPS62, GPS63, GPS64, GPS66, GPS67, GPS68, GPS69, GPS70, and GPS73.

Clock bias data include the following terrestrial stations with their given 4-character IDs: BRUX,\textsuperscript{*} CHPI, CIBG, DJIG, FTNA, GAMB, GLPS, GUAM, HARB, JFNG, JPLM, KOKV,\textsuperscript{*} KRGG, MAS1, MCM4, MDO1, NICO, NNOR,\textsuperscript{*} NYA2, OWMG, PALM, POL2, SANT, SGOC, STHL, STJ3,\textsuperscript{*} TIDB, VOIM, and YELL.\textsuperscript{*}
(\textsuperscript{*}Indicates H-maser clock.)

Since the H-maser clock at station KOUR is used as the reference clock for carrier phase data processing, clock bias data for KOUR are all implicitly zero, and are not provided in the submitted files. Data preparation for ELF analysis transformed the reference clock from KOUR to station KOKV by subtracting KOKV clock bias data from data for each satellite.

\subsection{Repeating geometry and common-mode errors}
\label{Sec:GPS-data-processing:repeat}

An aspect of the GPS system that we exploit is that, in the co-rotating ECEF reference frame of the global GPS station network, the satellite positions are designed to repeat approximately every sidereal day,~\cite{Agnew-repeat,Larson-repeat}, which is 1 day minus 236~s. The precise repeat time must account for the precession of the GPS orbits at $-14.4^{\circ}/$yr, caused by Earth's oblateness, and is a function of the orbit's eccentricity, inclination, and semi-major axis, corresponding to a repeat period of 1~day minus 245~s. This precession is slightly different for each satellite, resulting in a repeat period that is observed to vary in the range of 1~day minus $245\pm5~$s~\cite{Larson-repeat}.  The repeating orbit geometry causes systematic errors such as ground multipath reflections to also repeat, hence errors in the carrier phase measurement also tend to repeat.  Therefore, there is the potential to enhance signal-to-noise ratio by comparing  clock pseudo-frequency series from one repeat cycle to the next, as in common mode error subtraction techniques.  Such methods have been shown to reduce scatter in geodetic GPS positioning every second, resulting in accuracy at the few-mm level~\cite{Larson-repeat}; therefore, the signal-to-noise ratio in clock pseudo-frequency data should be similarly improved.  To confirm that we have repeating errors, we find that the autocorrelation of our own clock data peaks with a mean lag of 1~day minus 245~s, which corroborates orbit perturbation theory and previous observations~\cite{Larson-repeat}.

\subsection{Pseudo-frequency data}
\label{Sec:GPS-data-processing:Pseudo-frequency}


The Rb-locked local oscillator frequency is measured by a narrow-band circuit whose signal can be related to a sinusoidal wave~\cite{0.A.Howe0.W.Allan1981}. The output voltage of such a circuit can be expressed as
    \begin{align}
 V(t)=\left[V_0+\delta V(t)\right]\sin\left[2\pi \nu_\mathrm{clock} t+\Phi(t)\right]\,,\label{Eq:Voltage}
    \end{align}
where $V_0$ and $\nu_\mathrm{clock}$ are nominal peak voltage amplitude and the nominal clock frequency respectively; $\delta V(t)$ is the deviation from the nominal peak voltage and $\Phi(t)$ is the bias in clock phase~\cite{0.A.Howe0.W.Allan1981,Riley}. The clock phase bias can be caused by oscillator drift, technical noise, or an ELF signal. The clock bias in units of seconds can be written as a change of units,
    \begin{align}    
    T(t)=\frac{\Phi(t)}{2\pi \nu_\mathrm{clock}}\,. \label{Eq:Tbias}
    \end{align}
The instantaneous angular frequency $2 \pi \nu(t)$ is conventionally defined as the first derivative of the total phase; therefore from Eqs.~\eqref{Eq:Voltage} and \eqref{Eq:Tbias}, instantaneous frequency $\nu(t)$ is given by 
    \begin{align}
              \nu(t) =  \nu_\mathrm{clock}+\frac{1}{2\pi}\frac{d\Phi(t)}{dt} =  \nu_\mathrm{clock}+ \nu_\mathrm{clock}\frac{dT(t)}{dt}. \label{Eq:InstFreq}
    \end{align}
The clock's fractional frequency offset $y(t)$ is conventionally defined as
\begin{align}
    y(t) \equiv \frac{\nu(t)- \nu_\mathrm{clock}}{ \nu_\mathrm{clock}} = \frac{dT(t)}{dt}.
    \label{Eq:FFO}
\end{align}
Given sampled clock bias data, the pseudo-frequency is defined as
\begin{align}
    {T(t+\Delta_t)-T(t)}\approx y(t)\Delta_t, \label{Eq:PFreq}
\end{align}
where $\Delta_t$ is our data sampling interval.  Defined in this way, pseudo-frequency is in units of seconds. In our case where $\Delta_t=1\,\mathrm{s}$, pseudo-frequency can be conveniently interpreted as the fractional frequency offset (FFO).

By computing streams of pseudo-frequency data, our strategy is to seek evidence for GPS clock frequency excursions relative to a reference clock, which can then be analyzed in the context of theoretical ELF waveforms.

\subsection{Time series preparation}
\label{Sec:GPS-data-processing:Time-series-preparation}

We work with GPS satellite clock bias data for the day of the LIGO trigger: GW170817, detected on August~17, 2017 and the two prior days. Data for the days prior to the event are used to establish the characteristics of the background noise. The clock biases are referenced to a selected terrestrial clock. The discrete clock bias data stream $d^{(0)}_{a,j}$, sampled at every data epoch $j$, can be expressed for the $a^{\text{th}}$ satellite clock as
\begin{align}
   {d^{(0)}_{a,j}}=T_a(t_{j}) - T_r(t_j). \label{Eq:Clock-bias}
\end{align}
 The time-dependent clock biases of the satellite and reference clocks are denoted by $T_a(t_j)$ and $T_r(t_j)$ respectively. The reference time is fixed in the GPS data processing by setting $T_r(t_j)=0$ for all $j$.  It is shown explicitly here to remind us that data streams from all clocks will have an identical component of correlated noise arising from the reference clock.  Thus, correlated excursions in clock bias data streams may be a result of reference clock excursions from ideal time.  As a consequence, the quality of the reference clock imposes a lower bound on the detectable amplitude of ELF signals, if they exist.

Over the time window of a day, the Rb clock bias time series are dominated by a constant frequency drift plus random-walk noise. To assist statistical analysis, we whiten the data by first-order sequential differencing of the discrete clock biases~\cite{Roberts2017-GPS-DM},
\begin{align}
       {d^{(1)}_{a,j}} = {d^{(0)}_{a,j+1}-d^{(0)}_{a,j}}.\label{Eq:pseudo}
\end{align}

Considering that in our case, $\Delta_t=1\,\mathrm{s}$, and that our clock bias data is in units of seconds, the above expression computes the discretized pseudo-frequency~\eqref{Eq:PFreq}.  Thus, our pseudo-frequency data is a measure of FFO of the satellite clock relative to the reference clock. 
Here $j=\overline{1,N_t}$, where $N_t$ is the total number of pseudo-frequency data points.

Next, we remove the frequency drift for each satellite $a$ by subtracting the median pseudo-frequency data for that day:
\begin{align}
   \tilde{d}^{(1)}_{a,j}=d^{(1)}_{a,j}-\mathrm{Median}\left[   \{d^{(1)}_{a,j}\}_{j=1}^{N_t}\right]\,.
   \label{Eq:data}
\end{align}
We choose to use the median to avoid dependence on outliers or on the signal itself, if it exists.

Finally, we take a network median over $N_c$ $^{87}\mathrm{Rb}$ clocks as
\begin{align}
   \tilde{d}_{j}=\mathrm{Median}\left[   \{\tilde{d}^{(1)}_{a,j}\}_{a=1}^{N_c}\right].\label{Eq:Net_Med_Data}
\end{align}

In our search here, we consider $N_c=18$ $^{87}\mathrm{Rb}$ clocks from the GPS network.  Note that our analysis does not span day boundaries, because the original GPS carrier phase data are processed in daily batches, thus we can expect discontinuities in clock bias data at the day boundaries. As discussed in Sec.~\ref{Sec:GPS-data-processing:repeat}, the orbital geometry of the satellites repeats relative to an observer on Earth, leading to repeated multipath errors. As a result, the orbit repeat time in the ECEF reference frame is 1~day minus 245~s = $86,155\,\mathrm{s}$ for all satellites. This choice allows for the suppression of common-mode error while simplifying the analysis of network statistics.

We impose the requirement that only satellites with complete time-series data over the full observation window, three days in our case, are included in the analysis. This criterion simplifies the time-series processing and mitigates statistical heteroskedasticity arising from missing data. 

\section{Exotic low-mass fields}
\label{Sec:ELFs}

This section provides further details on parametrization of exotic low-mass fields (ELF) waveforms and their atomic clock signals.

 Massive fields satisfy the relativistic energy-momentum dispersion relation
  \begin{equation}
\omega(k)=\sqrt{(c k)^2+\Omega_c^2},
\label{Eq:dispersion}
\end{equation}
where the Compton frequency $\Omega_c = mc^2/\hbar$ depends on the ELF mass $m$.  The energy and momentum of the ELF quanta are given by $\varepsilon = \hbar \omega$ and $p=\hbar k$, where $k$ is the wave number.

An emitted ELF wavepacket can be described as a superposition of spherically-symmetric waves:
$\phi_k(r,t) = \frac{A_k}{r}\cos\prn{ k  r - \omega(k) t  + \theta_k},$ 
where $r$ is the radial coordinate, $A_k$, $\theta_k$, $k$, and $\omega$ are the ELF amplitudes, phases, wavevectors, and frequencies, respectively. The initial Gaussian wavepacket can be represented as a linear Fourier combination of these spherical waves. 
Individual components $\phi_k(r,t)$ propagate with different phase velocities $\omega(k)/k$. As the wavepacket propagates outward, its envelope changes over time exhibiting dispersion with the center of the envelope moving at the group velocity 

\begin{align}
    v_g = \frac{d\omega(k)}{dk}\Biggr\rvert_{k=k_0}\lesssim c\,.\label{Eq:vg}
\end{align} 

We assume that the ELF pulse is initially Gaussian, with central frequency
$\omega_0$ and initial pulse duration $\tau_0$, and is emitted from a
progenitor at distance $R$ from Earth. The resulting waveform is provided in
the main text.
 
The dimensionless quantity $\xi$ of the main text evaluates to
\begin{align}
    \xi=\frac{\Delta v_g t_s}{v_g \tau_0}\approx \left(\frac{1}{\omega_0\tau_0}\right) \frac{2\delta t}{\tau_0}\,,\label{Eq:xi}
\end{align}
where we the spread in the ELF group velocity $
    \Delta v_g= \Delta k\,{d^2\omega(k)}/{dk^2}|_{k=k_0}$,
with $\Delta k=|k-k_0|$ being the spread in the wave number.

The instantaneous ELF frequency is time-dependent, $\omega(t)  = d\theta(t)/dt$, exhibiting a frequency ``anti-chirp" at the sensor. The anti-chirp ramp is given by
\begin{equation}
 \frac{d\omega(t)}{dt}\approx\left\{
\begin{array}
[c]{cl}%
-\xi/\tau_0^2  \,, &  \text{$\xi\ll1$} \label{App:Eq:Signals}\\
-1/\tau \tau_0\,, &  \text{$\xi\gg1$}
\end{array}
\right. \,. 
\end{equation}
This reflects the qualitative fact that the more energetic (higher frequency) components have higher phase velocity $\omega/k$ thus arriving at the sensor earlier.

\subsection{Transient variation in fundamental constants}\label{Sec:Trans-Var}

ELFs can couple to atomic clocks through interactions that lead to the variation of fundamental constants.

The phenomenological interaction Lagrangian of the main text, leads to the effective redefinition of the fundamental constants (FCs). For clocks, the relevant FCs  are the
fermion masses and the fine structure constant,
\begin{align}
&m^{\mathrm{eff}}_f(\textbf{r},t)=m_f\left[1+\Gamma_f \hbar c \phi^2(\textbf{r},t)\right]\,\label{Eq:Fermion-mass-Redef},\\
    &\alpha^{\mathrm{eff}}(\textbf{r},t)=\alpha \left[1+\Gamma_{\alpha}\hbar c\phi^2(\textbf{r},t)\right]\,.\label{Eq:Fine-Structure-Const-Redef}
\end{align}

Variations in FCs would affect clock frequencies and, thereby, the apparent time as measured by the clock.
In general, the sensitivity of the oscillator resonance frequency $\nu$ to the FC variation $X$ can be quantified by the coefficient
\begin{align}
    \kappa_X=\frac{\partial\, \log \nu}{\partial \log X}\,.
    \label{Eq:kappaX}
\end{align}

The fractional clock frequency excursion due to quadratic interactions can be parameterized as (c.f.  main text)
\begin{align}
    s(t) = \frac{\delta\nu(t)}{\nu_{\mathrm{clock}}}=\Gamma_{\mathrm{eff}}\hbar c\phi(t)^2 \,,\label{Eq:ELF-sig}
\end{align}
where $\nu_{\mathrm{clock}}$ is the unperturbed (nominal) clock frequency, $\phi(t)$ is the ELF field at the sensor, and $\Gamma_{\mathrm{eff}}\equiv \sum_X K_X\Gamma_X$ is the effective coupling constant expressed in terms of the differential sensitivity coefficients $K_X$.

The sensitivity coefficients $K_X$ used in this analysis were derived in our
previous work~\cite{SenPfeRies2024-GPS.ELF}. In brief, our study effectively
compares rubidium ($^{87}\mathrm{Rb}$) microwave clocks aboard GPS satellites
with a hydrogen-maser reference clock. In the 1~Hz sampling regime relevant to
our data analysis, the differential sensitivity coefficients are independent
of the local oscillator. Using the sensitivity coefficients reported in
Refs.~\cite{Campbell2021,FlambaumEtAl2004,Roberts2018a}, the effective
coupling constant for the Rb–H clock comparison is~\cite{SenPfeRies2024-GPS.ELF}
\begin{align}
    \Gamma_{\mathrm{eff}}
    = 0.34\,\Gamma_{\alpha}
      + 0.081\,\Gamma_{m_q}\,,
    \label{Eq:GammaEff}
\end{align}
as quoted in the main text.

\section{Search for ELF signal}\label{Sec:ELF-search}

\subsection{Standard deviation analysis}

In the main text, we characterize the noise properties of the pseudo-frequency
data stream $\{\tilde{d}_{j}\}$ using the standard deviation computed over
finite-duration segments. The sequence $\{\tilde{d}_{j}\}$ is first partitioned
into $W$ windows, each containing $N_p = N_t/W$ samples. We denote the samples
in window $w$ by $\tilde{d}_{w,k} \equiv \tilde{d}_{j}$, where
$w = 1,\ldots,W$ labels the window and $k = 1,\ldots,N_p$ indexes the samples
within that window, such that $j = (w-1)N_p + k$. The standard deviation in the
$w$-th window is then defined as
\begin{align}
    \sigma_w =
    \sqrt{\frac{1}{N_p-1}
    \sum_{k=1}^{N_p}
    \left(
    \tilde{d}_{w,k}
    - \frac{1}{N_p}
    \sum_{k=1}^{N_p}
    \tilde{d}_{w,k}
    \right)^2 }\,.
\end{align}

In Fig.~\ref{Fig:STDEV}, the standard deviation $\sigma_w$ for each of the
three days is plotted as a function of the time $t_w$ associated with the
beginning of the $w$-th window, $t_w = w N_p \Delta_t$. The standard deviation
traces for August 15, 16, and 17, 2017 are largely comparable, indicating
similar levels of background variability across the three observation days.
This behavior is consistent with the expected daily repeating common-mode
structure of the constellation (see
Sec.~\ref{Sec:GPS-data-processing:repeat}). For August 15 and 16, the standard
deviation remains relatively stable throughout, with only small fluctuations.

\begin{figure}[ht!]
    \centering
    \includegraphics[width=1.0\columnwidth]{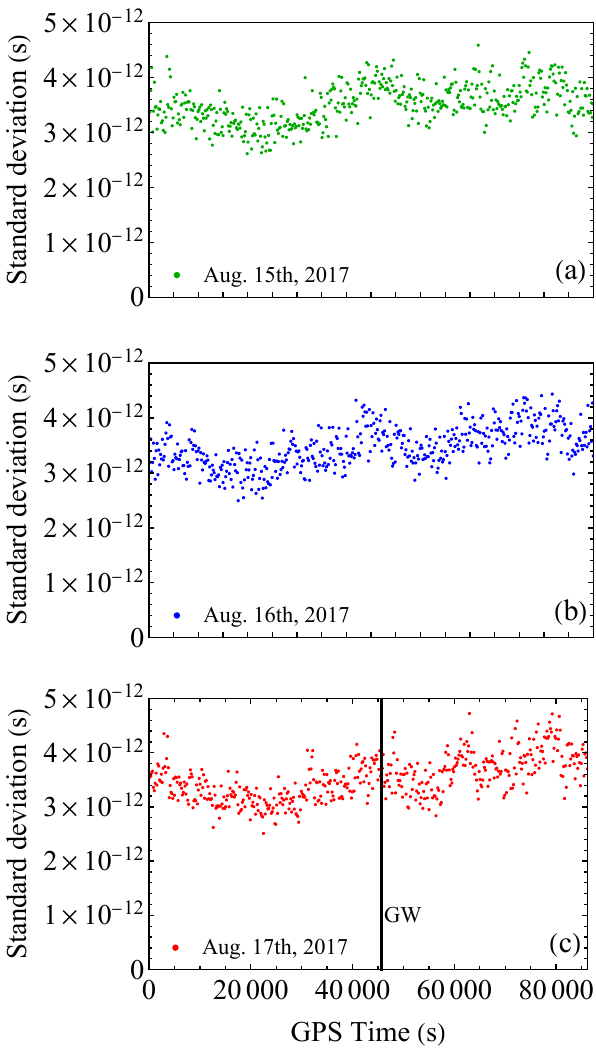}
    \caption{Standard deviation in a window, $\sigma_w$, of the network-median
    pseudo-frequency data~\eqref{Eq:Net_Med_Data} for the GPS network of
    $^{87}\mathrm{Rb}$ clocks on August 15, 16, and 17, 2017. The GW trigger at $45{,}682~\mathrm{s}$ from the start of August 17, 2017 (GPST),
    is indicated by the black vertical line. For each day, the data set is
    partitioned into $W = 498$ windows, each containing $N_p = 173$ data
    points, and $\sigma_w$ is computed within each window.}
    \label{Fig:STDEV}
\end{figure}

\subsection{Signal-to-noise ratio statistic}
\label{Sec:ELF-AmpStat}

Let the ELF signal~\eqref{Eq:ELF-sig} be expressed as $s(t_j)=hf(t_j)\equiv s_j$, where  $h$ is the signal amplitude given by
\begin{align}
    h=\left(\frac{\hbar c A_0^2}{R^2}\right)\Gamma_{\mathrm{eff}}\,,\label{Eq:b}
\end{align}
and $f(t_j)$ is the time-sampled template function 
\begin{align}
    f(t_j)=\exp\left[ -\frac{(t_j-t_{s})^2}{ \tau^2} \right] \cos^2\theta(t_j)\,.\label{Eq:fk}
\end{align}
Using notation of the main text, phase $\theta$  is
\begin{align}
   \theta(t_j)= & \omega_0 \left(t_j-t_\mathrm{GW}\right) 
   -\frac{1}{2\tau^2}\frac{\xi t_j}{t_s}\left(t_j-t_s\right)^2  + \nonumber \\
   &\frac{1}{2}\tan^{-1}\left(\frac{\xi t_j}{t_s} \right)+\theta' \,.\label{Eq:phase2}
\end{align}
The variable $t_j$ denotes the time since the emission of the ELF pulse. In our analysis, we adopt more practical shifted grid $t'_j=t_j-t_\mathrm{GW}$.
In this notation, the template becomes
\begin{align}
  f(t'_j)=\exp\left[  -\frac{(t'_j-\delta t)^2}{ \tau^2} \right] \cos^2\theta(t'_j)\,,\label{Eq:fkprime}  
\end{align}
with
\begin{align}
    \theta(t'_j)\approx \omega_0t'_j-\frac{\xi}{2}\left(\frac{t'_j +t_\mathrm{GW} }{\delta t+t_\mathrm{GW}}\right) \left( \frac{t'_j-\delta t}{\tau}\right)^2+\theta''\,,\label{Eq:theta-prime}
\end{align}
where $\theta''\equiv \theta'+\frac{1}{2}\tan^{-1}\xi$ and we used  $\tan^{-1}\left( \xi \frac{t_j}{t_s}\right) \approx \tan^{-1}\!\xi$.

As discussed in the main text, the data on the days preceding the merger are
well described by an approximately Gaussian distribution. We therefore model
the network-median data array $\{\tilde{d}_j\}$,
Eq.~\eqref{Eq:Net_Med_Data}, as the sum of normally distributed noise
$n_j \sim \mathcal{N}(\mu,\sigma^2)$ with mean $\mu$ and variance $\sigma^2$
and a possible signal component $s_j$, which in vector form reads
$\tilde{\boldsymbol{d}} = \boldsymbol{n} + \boldsymbol{s}$. The effective
noise variance $\sigma^2$ is estimated by averaging the variances obtained from
the two days of data preceding the event day. The mean of
$\{\tilde{d}_j\}$ is found to be approximately zero, so that the 
distribution of the ELF signal amplitude is given by
\begin{align}
   p\left(h|{\tilde{\mathbf{d}}}\right) 
        & \propto   \exp\left[-\frac{1}{2\hat{\sigma}_h^2} \left( h-\hat{h}\right)^2\right]\,, \label{Eq:Posterior}
\end{align}
where
\begin{align}
    &\hat{h}=\frac{\sum_{j=1}^N \tilde{d}_j f(t'_j)}{\sum_{j=1}^N f(t'_j)^2}\,,\label{Eq:hat-h}\\
    &\hat{\sigma}_h^2=\frac{\sigma^2}{{\sum_{j=1}^N f(t'_j)^2}} \,\label{Eq:sigma-h}.
\end{align}

The summations extend from the LIGO trigger to the end of August 17, 2017 in GPST scale. Hence, the first index $(j=1)$ in the primed time-scale ranges up to $N_\mathrm{GW}=t'_\mathrm{GW}/\Delta_t$ in the GPS day and the final index $(j=N)$ corresponds to $N_t$. Therefore, the total number of data points $N=N_t-N_\mathrm{GW}+1$. In practice, considering the exponential envelope of the signal, it is sufficient to sum over a window centered at $\delta t$ and spanning a few $\tau$. Based on Eq.~\eqref{Eq:Posterior}, the probability density function (PDF) of the amplitude estimator $\hat{h}$ is
\begin{align}
    p_h\left(\hat{h}|h\right)= \frac{1}{\hat{\sigma}_h\sqrt{2\pi}}\exp\left[-{\left( \hat{h}-h\right)^2}/({2\hat{\sigma}_h^2}) \right]\,.
\end{align}

The signal-to-noise ratio (SNR) statistic is defined as
\begin{align}
    \hat{\lambda} = \frac{|\hat{h}|}{\hat{\sigma}_h}\,.
    \label{Eq:SNR}
\end{align}
By standard error-propagation, the estimated uncertainty in the SNR statistic
$\hat{\lambda}$ is
$\hat{\sigma}_{\lambda}
= \sqrt{\left(\partial \hat{\lambda}/\partial \hat{h}\right)^2
\hat{\sigma}_{h}^{2}} = 1$.
Since $\hat{\lambda}\ge 0$ and $\hat{h}\sim\mathcal{N}(h,\hat{\sigma}_h^2)$,
the statistic in Eq.~\eqref{Eq:SNR} follows a folded-normal probability
distribution~\cite{Tsagris}, with PDF
\begin{align}
    p_{\hat{\lambda}}\!\left(\hat{\lambda}\,|\,\lambda\right)
    =& \frac{1}{\sqrt{2\pi}}
    \exp\!\left[-\frac{\left(\hat{\lambda}-\lambda\right)^2}{2}\right] + \nonumber \\
    &  \frac{1}{\sqrt{2\pi}}
    \exp\!\left[-\frac{\left(\hat{\lambda}+\lambda\right)^2}{2}\right]\,,
    \label{Eq:FoldedNormal}
\end{align}
where, in the frequentist sense, $\lambda$ denotes the (unknown) true value of
the statistic.

\subsection{Parameter space}
\label{Sec:Parameter-space}

The ELF template $f(t'_j)$ depends on three parameters,
$(\tau_0,\delta t,\omega_0)$, which we refer to collectively as a
\emph{triple}. In our search, we scan the ELF parameter space spanned by these
triples. In this section, we describe the sampling strategy and the
constraints applied to the parameter space.

Although the parameters $(\tau_0,\delta t,\omega_0)$ are continuous, we
discretize them for the numerical search. Since the time step of the data grid
is $\Delta_t$, the minimum value of the initial pulse duration $\tau_0$ is
$\tau_{0,\mathrm{min}}=\Delta_t$. The values of $\tau_0$ are then incremented
in steps of $\Delta_t$ up to $\tau_{0,\mathrm{max}} = N\Delta_t/4$, targeting
ELF signals whose temporal width lies well within the analysis window. The
delay parameter $\delta t$ is scanned on a grid with
$\delta t_{\mathrm{min}} = 2\Delta_t$ and
$\delta t_{\mathrm{max}} = (N-2)\Delta_t$, using a step size of $\Delta_t$.
This choice ensures that the leading and trailing edges of the ELF pulse are
properly sampled, which implies
\begin{align}
    2\tau \le \delta t \le N\Delta_t - 2\tau \,.
    \label{Eq:delta t-rel-1}
\end{align}
Note that the pulse durations at Earth $\tau$ and $\tau_0$ are related by
$\tau = \tau_0\sqrt{1+\xi^2}$.

Finally, we sample the ELF central frequency $\omega_0$ on a discrete grid.
The GPS data have a sampling rate of $1/\Delta_t$, so the minimum and maximum
values of the ELF central frequency are
$\omega_{0,\mathrm{min}} = 2\pi/(N\Delta_t)
\approx 1.6\times10^{-4}~\mathrm{rad/s}$
and
$\omega_{0,\mathrm{max}} = 2\pi/\Delta_t
\approx 6.3~\mathrm{rad/s}$, respectively.
We construct a logarithmically spaced set of values $\{\omega_{0,i}\}$ as
\begin{align}
    \omega_{0,i}
    = \omega_{0,\mathrm{min}}
      \left(N\right)^{i/(N-1)},
    \quad i = 0,\ldots,N-1 \, .
\end{align}
This sampling provides logarithmic coverage of $\omega_0$, consistent with the
energy axis used in the exclusion plots.

The three-dimensional parameter space spanned by the triples
$(\tau_0,\delta t,\omega_0)$ is further constrained by the condition
$\omega_0 \tau_0 \gg 1$. This requirement ensures that the ELF pulse remains
sufficiently sharp and that the parabolic approximation to the dispersion
relation~\cite{dailey2020ELF.Concept} remains valid. All combinations of parameter triples that satisfy
these selection criteria are referred to as \emph{valid triples}. Each valid
triple parametrizes a signal-template vector $\boldsymbol{f}^{\,i}$, whose
components are given by
\begin{align}
   f^i_{j}=\exp\left[  -\left( \frac{t'_j-\delta t^{i}}{\tau^i} \right)^2\right] \cos^2\theta^{i}(t'_j)\,,
\end{align}
where
\begin{align}
   &  \theta^i(t'_j)\approx \omega_0^it'_j-\frac{\xi^i}{2}\left(\frac{t'_j +t_\mathrm{GW} }{\delta t^i+t_\mathrm{GW}}\right) \left( \frac{t'_j-\delta t^i}{\tau^i}\right)^2+\theta''^i\,,\\
   &\theta''^i=\theta'+\frac{1}{2}\tan^{-1}\xi^i\,,\\
  & \xi^i=\left(\frac{1}{\omega_0^i\tau_0^i}\right) \frac{2\delta t^i}{\tau_0^i}\,, \\
  &\tau^i=\tau_0^i\sqrt{1+(\xi^i)^2}\,.
\end{align}
Here, the index $i = 1,\ldots,M$ enumerates the templates, where $M$ is the
total number of templates (i.e., valid triples). By scanning the ELF parameter
space, we obtain $M \approx 6\times 10^{12}$. The collection of all
$\boldsymbol{f}^{\,i}$ constitutes the template bank used in our ELF search.

\subsection{Search methodology}

The search is performed by scanning over the template bank and evaluating the
SNR statistic~\eqref{Eq:SNR} for each template, in order to assess whether any
template exhibits substantial statistical agreement with the network-median
pseudo-frequency data~\eqref{Eq:Net_Med_Data}, i.e., the stream
$\{\tilde{d}_j\}$. For each template, we draw values of the random phase
$\theta'$ from a uniform distribution and numerically maximize the
SNR~\eqref{Eq:SNR} with respect to $\theta'$.

We adopt a frequentist hypothesis-testing framework~\cite{RomanoCornish2017}
for the ELF search. In this approach, evidence for a signal is assessed by
testing a null hypothesis $H_0$ corresponding to the absence of a signal in the
data. Under the assumption that $H_0$ is true, the sampling
distribution~\eqref{Eq:FoldedNormal} of $\hat{\lambda}$ reduces to
\begin{align}
    p_{\hat{\lambda}}\!\left(\hat{\lambda}\,|\,\lambda = 0, H_0\right)
    = \sqrt{\frac{2}{\pi}}
      \exp\!\left(-\frac{\hat{\lambda}^2}{2}\right)\,.
    \label{Eq:PDF_H0}
\end{align}

If the observed value of the SNR statistic, $\hat{\lambda}_{\mathrm{obs}}$,
lies sufficiently far in the upper tail of the sampling distribution, the data
are no longer consistent with $H_0$, and $H_0$ may be rejected in favor of the
alternative hypothesis that a signal is present in the data. A detection
threshold $Z_1^*$ is defined in terms of a chosen false-positive probability
$q_1$~\cite{RomanoCornish2017} as
\begin{align}
&\nonumber \int_{Z^*_1}^\infty   p_{\hat{\lambda}}\left(\hat{\lambda}|\lambda=0,H_0\right) d\hat{\lambda}=q_1\,,\\
\Rightarrow~ &Z^*_1=\sqrt{2}\erf^{-1}\left(1-q_1\right)\,.\label{Eq:Thr}
\end{align}

Fixing the conventional SNR threshold \(Z_1^* = 3\) in
Eq.~\eqref{Eq:Thr} yields a corresponding false-alarm probability of
\(q_1 = 0.0027\). This threshold, however, is valid only for a search
involving a single template, which is the reason for the subscript ``1'' in the
notation.

As the number of templates $M$ increases, the probability that random noise
produces a spurious template match, and hence a false positive, increases as
well. In this case, the SNR threshold must be redefined. For each template
$i=1,\ldots,M$, we compute the SNR statistic
$\hat{\lambda}^i$~\eqref{Eq:SNR}, and we define
$\hat{\lambda}$ as the maximum value over all templates,
i.e.\ $\hat{\lambda}=\max\{\hat{\lambda}^i\}$. The sampling distribution of
$\hat{\lambda}$ is governed by the correlation structure of the template bank
and may be described by a multivariate normal model~\cite{Tyler_Andrei_2022}
with bank correlation matrix $\boldsymbol{\Sigma}$, whose elements are
\begin{align}
    \Sigma^{i,k}
    = \frac{\boldsymbol{f}^{\,i}\!\cdot\!\boldsymbol{f}^{\,k}}
    {\sqrt{(\boldsymbol{f}^{\,i}\!\cdot\!\boldsymbol{f}^{\,i})
           (\boldsymbol{f}^{\,k}\!\cdot\!\boldsymbol{f}^{\,k})}}\,,
    \label{Eq:TempCorrMatrix}
\end{align}
where the superscripts $i$ and $k$ enumerate the templates defined in
Sec.~\ref{Sec:Parameter-space}. In our search, the template bank is found to be
nearly orthogonal, so that $\boldsymbol{\Sigma}\approx\boldsymbol{I}$, where $\boldsymbol{I}$ is the identity matrix. Then the threshold $Z = Z_M^*$ satisfies
\begin{align}
    C_M(Z_M^* \,|\, \boldsymbol{I}) &= 1 - q_1 \,,
    \nonumber\\
    \Rightarrow\quad
    Z_M^* &\approx
    \sqrt{2}\,\erf^{-1}\!\left(1-\frac{q_1}{M}\right)\,,
    \label{Eq:Zstar}
\end{align}
where $C_M(Z)$ denotes the cumulative distribution function of $Z$.  For
$M\approx6\times10^{12}$ and a fully orthogonal template bank, we obtain
$Z_M^*\approx 8$. For a template bank that is only nearly orthogonal, the
corresponding threshold will be somewhat lower~\cite{Tyler_Andrei_2022}.

In a complete template-based search, the ELF parameter space spanned by the
triples $(\tau_0,\delta t,\omega_0)$ would be sampled exhaustively. However,
for $M\approx 6\times 10^{12}$ templates, such a search is computationally
prohibitive. To mitigate this, we implemented a sparse-scanning strategy in
which the three-dimensional parameter space is explored using nested loops
over the triples, and the SNR is evaluated for every billionth unique triple.
This approach provides uniform coverage of the parameter space while
substantially reducing computational cost. In this sparse search, no SNR value
exceeded even the single-template threshold $Z_1^*=3$, indicating no candidate
ELF events within the sampled parameter set.

While the sparse search has not revealed any ELF signal, we notice that Fig.~\ref{Fig:STDEV}(c) shows fluctuations of duration~$\lesssim$~a few hundred seconds. To verify the absence of the ELF signal, we use the dense search strategy where we scan over all possible values of $\omega_0~\text{and}~\delta t$ while limiting {$\tau_0\le 500\,\mathrm{s}$}. The maximum SNR from this search is $\approx 3.8$ for $\tau_0=3~\mathrm{s}, \delta t = 30,443~\mathrm{s}, \omega_0=1.71~\mathrm{rad/s}$. 
From this dense search we found no candidate events exceeding $Z^*_M=8$.

\subsection{Setting limits}\label{Sec:Limits}
\begin{figure}[ht!]
    \centering  \includegraphics[width=1.0\columnwidth]{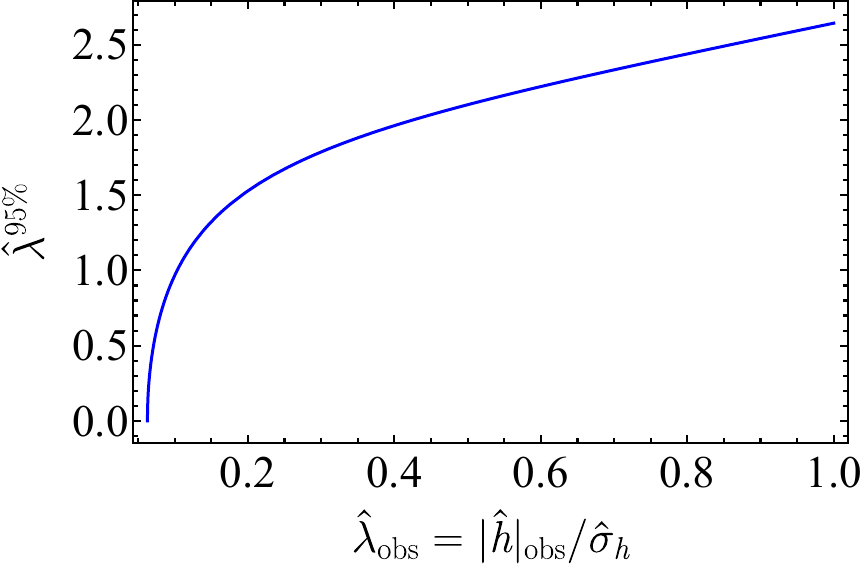}
    \caption{95\% confidence limit on the SNR $\hat{\lambda}$ as a function of observed SNR, $\hat{\lambda}_{\mathrm{obs}}$. Here, $|\hat{h}|_{\mathrm{obs}}$ denotes the absolute value of the estimated or observed ELF amplitude~\eqref{Eq:hat-h}, with an uncertainty of $\hat{\sigma}_h$.}\label{Fig:ULPlot}
 \end{figure}

In the absence of a detection, a statistical bound or upper limit (UL) is placed on the statistic $\hat{\lambda}$ as
\begin{align}
\mathbb P\left(\hat{\lambda}\ge\hat{\lambda}_{\mathrm{obs}}|\lambda=\hat{\lambda}^{\mathrm{UL}\%}\right)=\mathrm{UL} \,,\label{Eq:UL}
\end{align}
where $\mathbb P (\hat{\lambda})=\int_{a}^{b}p_{\hat{\lambda}}d\hat{\lambda}$ denotes the probability of finding $\hat{\lambda}$ between $a$ and $b$. With a 95\% limit on the SNR statistic, i.e. for $\mathrm{UL}=0.95$, Eq.~\eqref{Eq:UL} becomes
\begin{align}
     & \int_{\hat{\lambda}_{\mathrm{obs}}}^{\infty}  p_{\hat{\lambda}}\left(\hat{\lambda}|\lambda=\hat{\lambda}^{95\%}\right) d\hat{\lambda}=0.95\,,\label{Eq:UL95}
\end{align}
which is simplified by using Eq.~\eqref{Eq:FoldedNormal}  to 
\begin{align}
    \mathrm{erf}\left(\frac{\hat{\lambda}_{\mathrm{obs}}-\hat{\lambda}^{95\%}}{\sqrt{2}}\right) + \mathrm{erf}\left( \frac{\hat{\lambda}_{\mathrm{obs}}+\hat{\lambda}^{95\%}}{\sqrt{2}}\right) = 0.1\,.\label{Eq:UL95_Eq}
\end{align}
Solution mapping $\hat{\lambda}_\mathrm{obs}$ into $\hat{\lambda}^{95\%}$ is shown in Fig.~(\ref{Fig:ULPlot}).  

This procedure yields a 95\% upper limit on the ELF signal amplitude
$|\hat{h}|$ based on Eq.~\eqref{Eq:SNR}, namely
\begin{align}
    |\hat{h}|^{95\%}
    = \hat{\sigma}_h\,\hat{\lambda}^{95\%}\,,
\end{align}
which in turn implies a 95\% upper limit on the effective coupling constant
$|\Gamma_{\mathrm{eff}}|$. From Eq.~\eqref{Eq:b}, this is given by
\begin{align}
    |\Gamma_{\mathrm{eff}}|^{95\%}
    = \left(\frac{R^2}{\hbar c A_0^2}\right)
      |\hat{h}|^{95\%}\,.
\end{align}

\begin{figure}[ht!]
    \centering  \includegraphics[width=1.0\columnwidth]{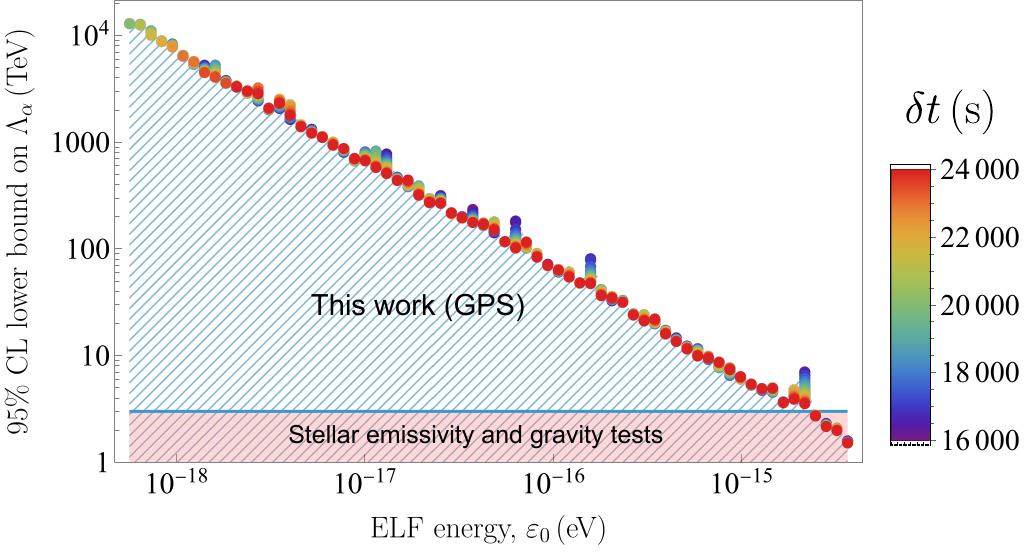}
    \caption{
    Shown are 95\% confidence-level (CL) lower bounds on the energy scale $\Lambda_\alpha$
governing quadratic couplings of ultralight scalar fields causing variations in fine-structure constant. The exclusion region for GPS atomic clocks (blue shaded area) is compared to that set by stellar emissivity and gravity tests~\cite{OlivePospelov} (pink shaded region).  This is computed for August 17, 2017 data stream. Here, the range of $\delta t$ is based on Eq.~\eqref{Eq:delta t-rel-1}. This $\delta t$  range is sampled every $500\,\mathrm{s}$.}\label{Fig:LimPlot1}
\end{figure}

Using Eq.~\eqref{Eq:GammaEff}, and assuming that the contribution from
variations of the fine-structure constant $\alpha$ dominates, the 95\% upper
limit on $|\Gamma_{\alpha}|$ is
\begin{align}
    |\Gamma_{\alpha}|^{95\%}
    \approx
    \frac{|\Gamma_{\mathrm{eff}}|^{95\%}}{0.34}\,,
    \label{Eq:Coup_Lim}
\end{align}
which may equivalently be expressed as a 95\% lower bound on the associated interaction
energy scale~\cite{dailey2020ELF.Concept},
\begin{align}
    \Lambda_{\alpha}^{95\%}
    = \frac{1}{\sqrt{|\Gamma_{\alpha}|^{95\%}}}\,.
\end{align}

Figure~4 of the main text and Fig.~\ref{Fig:LimPlot1} show the exclusion
region for the energy scale $\Lambda_{\alpha}^{95\%}$ as a function of the ELF
energy $\varepsilon_0$, evaluated over different ranges of $\delta t$ and
$\tau_0$. The blue shaded region is excluded by our ELF search. The most
stringent existing constraints, $\Lambda_{\alpha}\lesssim 3~\mathrm{TeV}$, are
set by limits on the thermal emission rate from supernova cores~\cite{OlivePospelov}.
Ref.~\cite{OlivePospelov} also examines complementary constraints derived from
precision tests of the gravitational force.

\end{document}